\documentclass[twocolumn]{webofc}
\usepackage[varg]{txfonts}   
\begin{document}
\title{Gravitational wave signatures of phase transition from hadronic to quark matter in isolated neutron stars and binaries}
%
%

\author{\firstname{Ritam} \lastname{Mallick}\inst{1}\fnsep\thanks{\email{mallick@iiserb.ac.in}}}

\institute{Department of Physics, Indian Institute of Science Education and Research Bhopal, India}

\abstract{%
The fundamental constituent of matter at high temperature and density has intrigued physicists for quite some time. Recent results from heavy-ion colliders have enriched the Quantum Chromodynamics phase diagram at high temperatures and low baryon density. However, the phase at low temperatures and finite (mostly intermediate) baryon density remain unexplored. Theoretical Quantum Chromodynamics calculation predicts phase transition from hadronic matter to quark matter at such densities. Presently, the best laboratories available to probe such densities lie at the core of neutron stars. Recent results of how such phase transition signatures can be probed using gravitational waves both in isolated neutron stars and neutron star in binaries. The isolated neutron star would probe the very low-temperature regime, whereas neutron stars in binaries would probe finite baryon density in the intermediate temperature regime. We would also discuss whether the gravitational wave signature of such phase transition is unique and the detector specification needed to detect such signals.
}
\maketitle
\section{Introduction}
\label{intro}

The quest to understand matter properties at high density and temperature has inspired physics for about a decade. Extensive collider experiments were designed and set up to probe matter properties at high temperatures and low densities. Recent results from collider experiments suggest that there is a crossover phase transition (PT) from hadronic matter (HM) to quark-gluon plasma at high temperatures \cite{shuryak}. The study of matter at high density and temperature is called quantum chromodynamics (QCD). QCD also predicts PT from nuclear matter to quark matter is also likely at high density and is assumed to be a first-order PT. However, creating laboratories to probe matter at high density is challenging, which is still not achieved in earth-based experiments. Another way of having some knowledge of matter properties at high density is to have reliable theoretical calculations. Accurate lattice calculation of QCD exists for zero or low-density matter (finite temperature), but again, for finite density, such calculations fail \cite{lattice_qcd,Goy_2017}. There are also some calculations for the matter at asymptotic high densities (perturbative QCD \cite{pqcd}), but there is not much progress for intermediate densities. This intermediate density range is particularly interesting because it is expected that the PT to HM to quark matter (QM) occurs at such a density regime.

The only laboratory to test the theory of strong interaction (QCD) at a finite density regime is at the cores of neutron stars (NSs). NSs are very compact objects with a mass of about $2$ solar mass and a radius of about $10-15$ km \cite{glen}. Theoretical models predict that at the cores of NSs, the density can be as high as $4-6$ times nuclear saturation density. Such density regimes are the most likely place where the PT from HM to QM can take place. However, the core of the NSs are not visible to us, and we have to model them accurately and then match the results with observations. The observational counterpart of NSs are pulsars, and recently improved astrophysical observation has provided an accurate measurement of NS mass \cite{demorest,Antoniodis_2013,Cromartie_2020}. Recently NICER mission has been launched, which has been able to measure the radius with a certain degree of accuracy \cite{Miller_2021,Riley_2021}.

On the other hand, an unexpected constraint on the matter properties also appeared from the gravitational wave (GW) observation of GW170817. The GW originated from the merging of two binary NSs \cite{GW170817} and was observed not only in LIGO but also in other electromagnetic and neutrino detectors. The GW observation constrains the tidal deformability of the NSs, which is governed by the star equation of state (EoS). Therefore, it also constrained the EoS to a certain degree \cite{Hinderer_2010,Most_2018,Annala_2018,Zhang_2018,Nandi_2021}. Therefore, the measurement of massive pulsars and the tidal constrain of the star have put severe constrain on the EoS. However, the ambiguity on the matter properties (HM or QM) at NS cores and possible signals of PT remains elusive.

It should be understood that the GW signal observed from GW170817 was from the pre-merger phase. However, the post-merger physics of NS mergers is expected to provide much more insight into the problem of PT. After the merger, there can be several outcomes depending on the mass of the merging NSs \cite{Weih_2020,Sarin_2021}. If the mergers are two massive NSs, they can easily collapse into a black hole (BH) immediately after the merging. However, suppose the mergers are of intermediate mass. In that case, they can either be for a hypermassive star or a supramassive star which can either collapse to a BH or lose its angular momentum to form a stable single NS. However, one thing is sure: after the two stars merge, the density of the merger products rises considerably. At such densities, it is most likely that there can be a PT from HM to QM, and such PT can have unique signatures \cite{Most_2019}. Simulation results also hint that there can be unique high-frequency peaks in the spectrum which correspond to the definite properties of the EoS \cite{bauswein_2017,gth,gw_ana,ppeos}.

The exact density where the PT line exists is not very definite. Therefore, PT can even be expected to happen in cold isolated NSs \cite{mallick-prc74,drago_2007,mallick-apj1,mallick-apj2,mallick-mnras-slow}. This can happen when the central density of massive NSs crosses the critical density for which HM is stable. NS's central density rises as it slows down after birth \cite{prasad-mnras}. This article will first study the condition under which the PT takes place in isolated NSs and how it dynamically evolves. Next, we will study the PT scenario in binary neutron star mergers (BNSM). We will also discuss the observational signatures obtained from such PT in isolated and binary NSs.

\section{PT in isolated NSs}

There are a plethora of guesses about how the PT seeds at the centre of the star. The earlier work \cite{alcock,glen} considers the seeding can be due to accretion, spin-down, cooling or seeding by a quark nugget. However, there is no detailed analysis of how this seeding happens. Assuming a 2-step PT \cite{mallick-prc74, mallick-apj1, mallick-mnras-slow} where first the deconfinement of nuclear matter to 2-flavour quark matter happens which is followed by the weak conversion of excess of down quarks to strange quarks. The first confinement transition is assumed to be a shock-induced PT. This happens when the star's centre attains a density above the critical density for which the quark matter is stable. 

To study the PT at the NS interior, one needs to have some model for HM and QM at neutron star cores. To describe HM EoS, we usually involve relativistic mean field theory EoS involving interactions between hadrons mediated by mesons \cite{serot,ellis,horowitz,typel}. To describe QM EoS, we usually involve modified MIT bag model EoS where interactions are described with a generalized parameter \cite{chodos,alford,weissenborn}. It is sometimes challenging to perform simulations with tabulated EoS; therefore, while performing simulations, we have used piece-wise polytropic fit of tabulated EoS \cite{mallick-apj1,haque-bnsm}.

\subsection{Seeding of quark matter at the core by spin down}

After the supernova (SN) explosion, the proto-neutron star is hot and has differential and angular rotations. As it cools, the differential rotation ceases, and the angular rotation slows down. Conserving angular momentum, the angular velocity of NSs is close to that of Keplerian velocity just after their birth. As the NS slows down (due to loss of energy and angular momentum), the central density of the star rises. Moreover, at some stage of their life, it slows down to such a value that the central density rises above the critical density below which the HM is stable \cite{mallick-prc-multi,prasad-mnras}. As soon as this happens, a large chunk of matter at the core of the NS converts to QM. If the process is exothermic and the density of the quark core is greater than its adjoined HM, a density discontinuity is generated at the core. The discontinuity can give rise to shock discontinuity, and a shock wave can propagate from the core to the star's outer surface. If the HM to QM combustion is exothermic, this gives rise to a shock-induced PT. The PT is usually a two-step process: the first step occurs when the shock propagates outwards and deconfines HM to 2-flavour QM. As the 2-flavour matter is meta-stable, it further decays to 3-flavour QM with the generation of strange quarks; however, this combustion front is much slower and follows the deconfinement front \cite{mallick-prc74,mallick-prc76}.

Assuming that the baryonic mass of the star remains constant throughout the NSs life, there is no mass ejection even during the PT process. The main idea of spin-down-induced PT is that the star's central density crosses the critical density at some point in its life. Therefore, for very low-mass stars, the central density of the star never rises beyond the critical density, and they remain an NS throughout their life. Intermediate mass stars are expected to have the core density rising beyond critical density at some point in their life, and they are expected to have PT in their lifetime. Relatively heavier stars can have a quark core at their centre after birth. The QM EoS at the star's centre is usually assumed to be softer than the HM EoS. Therefore as the QM core appears at the star centre, the stars become more compact (its radius changes). However, for massive stars, if the quark core at the centre of the star is large enough, then the matter pressure may not be able to balance the inwards gravitational collapse, and the star may collapse to a BH. All such situations are likely to occur for a spin-down-induced PT at NS cores; however, the exact critical density and the mass range for such a situation depend strongly on the EoS.

The highly deformed star during birth settles as it grows old and slows down. The star becomes less elliptic and its compactness also increases. This results in a change in the star's moment of inertia (MOI). However, when the quark matter first appears, the MOI changes significantly. Also, as the star slows down, the quark content of the star increases. Slowly rotating is expected to have larger quark cores than faster rotators. As the quark core grows, the star's radius shrinks, and the star becomes more compact. A compact star with a quark core and hadronic outer surface is known as a hybrid star (HS).

If we assume a vacuum dipole model for an NS and if the star loses its rotational energy by electromagnetic emission from its magnetic axis, we can calculate the time needed by a star to slow down and the density to cross the critical value for PT to happen. 
The NSs period can its derivative is given by \cite{Abdikamalov},

\begin{equation}
P=\sqrt{P_{0}^{2}+2 A t},
\end{equation}

\begin{equation}
\dot{P}=\frac{A}{\sqrt{P_{0}^{2}+2At}},
\end{equation}

where $P_{0}$ is its period at birth, $t$ is the age, and the quantity A is given by $A=\frac{8 \pi^{2} B_{s}^{2} R^{6}}{3 c^{3} I}$. The surface magnetic field is $B_{s}=B_{0} \sin(\beta)$, the asymptotic value $B_{0}$ and $\beta$ is the tilt axis which is assumed to remain constant. 
It can be easily found \cite{mnras-prasad} that the rate of slowing down of the star is faster for the massive star than for low-mass stars. Also, the slowdown is faster if the star's magnetic field is large. Depending on the star's mass and magnetic field, the star can take from months to thousands of years to slow down considerably and cross the critical density for PT.
 
\subsection{1st step: deconfinement transition}

Once the quark seed is generated at the star centre, it creates a discontinuity in the thermodynamic quantities at the quark hadron boundary. If the discontinuity is large enough, then the shock wave propagates outwards. We assume that as the shock wave travels it deconfines hadronic matter to 2-flavour QM. The distance this shock wave will travel will depend on the corresponding EoS of HM and QM. The dynamic evolution of the shock wave is studied using a general relativistic hydrodynamic equation. For simplicity, we have assumed the star to be spherically symmetric and used GR1D code to study the shock propagation, and the deconfinement \cite{mallick-apj1}. 

The PT is modelled similarly to that of the sod-type Riemann problem. On one side, we have HM; on the other, we have QM. The position of the shock discontinuity is located at every step, and the deconfinement happens at the shock boundary. Although the thermodynamic variables are discontinuous at the shock or deconfinement front, the energy-momentum and mass flux are conserved. This is repeated for every time step of the dynamic evolution. We have also assumed that the baryonic matter of the star remains constant, and no matter is ejected out as the shock travels outwards. It is seen that the shock strength decreases as it propagates out. It is because as we go to the outer regions, the matter density decreases and the difference in the thermodynamic variables across the front. On the other hand, the shock velocity increases as we go outwards. The shocks take a few tens of microseconds to travel across the star. The timescale for the deconfinement transition is quite significant because there is no other process that fits this timescale for NSs. This can have a significant outcome for the observation features of this deconfinement process.

However, if we want GW emission, we must have a non-zero quadrupole moment of the star whose double derivative must not vanish. A spherically symmetric star is insufficient, and we need an axis-symmetric star. Therefore, we extend the shock-induced confinement calculation for a rotating star. The rotating star is generated from the rotating neutron star code, and we have modified the GR1D code to accommodate a 2-d simulation of the general relativistic hydrodynamic equations. Although the shock discontinuity travels throughout the star, the deconfinement transition happens till the critical point.

The propagation of the shock discontinuity alters the internal composition of the star, and it happens very fast (in comparison to the star's rotational velocity). It changes the quadrupole moment of the star. The quadrupole moment of the star changes continuously as the discontinuity moves outwards; however, as the deconfinement transition ends, there is a sudden change in the star's quadrupole moment (when the deconfinement transition ends). It is critical for generating the GW, and the wave's strain depends very strongly on this sharp change in the quadrupole moment.

The GW strain for an axis-symmetric star is given by \cite{zwerger,Dimmelmeier}
\begin{equation}
h_{\theta \theta}^{TT} =\frac{1}{8} \sqrt{\frac{15}{\pi}} \sin^{2}\theta \frac{A_{20}^{E2}}{r},
\end{equation}
\begin{equation}
h_{\phi \phi}^{TT} =-h_{\theta \theta}^{TT}=h_{+}
\end{equation}
where $ \theta$ is the angle between the symmetry axis and the observer's line of sight. $A_{20}^{E2}$ is described as 
\begin{equation}
A_{20}^{E2}= \frac{d^{2}}{dt^{2}} \left( k \int \rho \left( \frac{3}{2} z^{2}-\frac{1}{2} \right)r^{4} dr dz \right)
\end{equation}
with 
$z=\cos{\theta}$ and $k=\frac{16 \pi^{3/2}}{\sqrt{15}}$.

The change in the quadrupole moment is carried out for every step, the hydrodynamic equation is solved, and the above formula is used to calculate the GW strain. A star which is rotating between $50-100$ Hz, the GW amplitude (strain) is around $10^{-21}- 10^{-22}$ for an NS at a distance of $10$ Mpc \cite{mallick-apj2}. The GW signal is a burst-type signal and is well within the present detector capability. However, the only problem with detecting such bursts type signals is that the power spectrum peak is around $100$ of kHz, which is much beyond the detector capabilities of present-day detectors.

\subsection{2nd step: weak combustion}

As the shock wave passes, it deconfines matter, and now the matter is in a metastable state having only up and down quarks (excess of up quarks as neutron matter has an excess of neutron than a proton). The appearance of strange quarks lowers the average energy per quark, and the matter settles into a stable equilibrium. The conversion of 2-flavour matter to 3-flavour matter is usually an exothermic process, and the combustion can happen readily inside the star. Thus after the first wave, a second front is generated at the star core, which converts 2-flavour matter to 3-flavour matter. The weak interaction process governs this combustion. As weak interaction is much slower than the strong deconfinement process, the second front always lags the first front \cite{mallick-mnras-slow}.

The second front separates the 2-flavour matter and 3-flavour matter. At a particular radial point inside the star at the front boundary, excess down quarks are converted to strange quarks. However, there is also diffusion across the boundary. The conversion stops when the down quark chemical potential becomes equal to that of the strange quark chemical potential. Then the front moves outward into the 2-flavour region. We have derived a differential equation with a parameter to analyze the process. The parameter `$a$' is function of radial coordinate $r$ taking center of star as the origin,
\begin{equation}\label{a}
a(r)=\frac{n^{2f}_{k} (r)-n^{3f}_{k} (r)}{2n_b(r)},
\end{equation}
where $n_k=\frac{1}{2} (n_d-n_s)$. Superscripts 2f and 3f denote the  2-flavour and 3-flavour matter, respectively. $n^{3f}_k$ is asymptotically the value of $n_k$ in the 3-flavour equilibrated matter and approximately close to $0$, the minimum value of $a$. For purely 2-flavour matter, $n_k=n^{2f}_d/2=n^{2f}$ due to absence of $s$ quarks in this  region $a \approx \frac{n^{2f}}{n_{b}} \approx 1$ which is the maximum possible value of $a$. Therefore, $a$ lies in the range between $0$ to $1$. The differential equation has two components, the decay rate and the diffusion coefficient, which are the function of quark chemical potential and temperature. The differential gets a correct asymptotic solution for the right value of the front velocity. It is also seen that the velocity of the weak combustion front decreases with a temperature rise.

Once the combustion velocity is obtained, we follow the same procedure by which we calculate the GW strain for the deconfinement process. The velocity of the front lies in the range of about $10^{-2}-10^{-4}$c. Therefore, the time taken by the weak combustion front to convert 2-flavour matter to 3-flavour matter is about a few milliseconds to tens of milliseconds. The GW strain ranges between $10^{-22}-10^{-25}$ for a star at a distance of about $10$ Mpc depending on the star's temperature. However, the power spectrum peak frequency is around a few $ 100$ Hz \cite{mallick-mnras-slow}. Although the GW strain is challenging to detect with present-day detectors, the frequency is well within detection capability.

\subsection{Signals associated with PT in isolated NSs}

The PT has two distinct bursts type GW signal, as is clear from the previous two sections. As they are burst-type type signals, they are not easy to capture, and unless we go onto the next generation of detectors, it is difficult to capture such signals. The first burst signal has a strong amplitude, but the frequency is challenging, whereas the second signal's frequency is well within detector capability. However, the strain might be difficult to detect with present detectors. However, along with the GW signals, PT can have other electromagnetic and neutrino signals. During the second process, a massive amount of neutrino-antineutrino pairs can be generated during the weak decay process. The neutrino-antineutrino pairs can annihilate and deposit a huge amount of energy at the star's surface. Depending on the star's mass, the energy can be of the order of $10^{49}-10^{50}$ ergs at timescales of tens of milliseconds. The energy budget is of the same order as that of short gamma-ray bursts \cite{mallick-prc-multi}.

As the star loses such a huge amount of energy and its internal composition and structure change, its tilt angle (angle between the star's symmetry axis and magnetic axis) changes rapidly. Depending on the star mass, the change in the star tilt angle can be between $1^o - 12^o$. The evolution of the tilt angle can have the sudden appearance (or disappearance) of the pulse signals from a pulsar. Usually, a rotating NS also emits continuous GW, which depends strongly on the tilt angle of the star. Therefore, a change in the tilt angle can significantly impact the continuous GW emission of the star \cite{mallick-prc-multi}.

\section{PT in binary NS mergers}

As discussed earlier, although PT in isolated NS is possible, PT is more likely to happen during BNSM, where the density of the merger product rises to large values. There has been quite a number of works which discusses the possible post-merger signals resulting from such PT at the core of the merger product \cite{most-rezzolla_2019,weih-rezzolla_2020,bauswein_prl_2019,tootle-ecker-rezzolla_2022}. The post-merger signals for PT depend strongly on the EoS of the corresponding HM and QM but also on the critical density above which the quarks start appearing in the system (also known as the onset point).

To set up the simulation, we constructed the initial data using the \textsc{Lorene} code~\cite{lorene2}. We constructed two NSs using the code with an initial physical separation of $40$ km. Next, we evolve the binary system (solving the hydrodynamic equation along with the geometric equations) using \textsc{Einstein Toolkit}~\cite{ET2,ETextra1,ETextra2,ETextra3} with \textsc{McLachlan}~\cite{mclachlan1,mclachlan2} (implementation of spacetime evolution) and \textsc{IllinoisGRMHD}~\cite{illinois1,illinois2} (GRMHD solver). The extraction of the gravitational waveform is done by calculating the Weyl scalar. The mode we analyze is the $l=m=2$ of the strain $h$ at a distance of $100$ Mpc. The merger time is the point where 
$|h^{22}|$ is maximum. Further, we have extracted the instantaneous frequency and also the Power Spectral Density (PSD) of the GW amplitude.

For the BNSM, we have done three simulations with equal mass binaries and one with unequal mass binary. The hybrid EoS has HM at low-density mixed phase region at intermediate density, and at high density, we have pure QM. The onset point of QM (where QM first appears in the system) has been varied to check how this affects the simulation and the GW signals from the post-merger product. For equal mass binaries, when two small NS merges, the initial star is composed entirely of HM. After the merging, the density of the merger product rises and depending on the onset point; quark matter can appear in the system. The exciting thing is that when small stars merge, the merger product is an NSs (does not collapse to a BH), whether quarks appear in the system or not. However, the GW signals differ in amplitude and phase depending on whether quarks are present in the resultant hypermassive star (HYS). The phase difference increases if the onset point happens at low density. However, if two more massive star merges $\sim 1.4~M_\odot$ and the onset point is at low density, even the merging stars have a quark core. Therefore, with such a star, the merger product collapses to a BH shortly after merging; however, if QM does not appear or appears at large densities, then it does not usually collapse to a BH (at least it is stable for some time). However, if two large NS merges ($\sim 1.6~M_\odot$), the ultimate of the merger product is always a BH irrespective of the EoS (HM or QM) \cite{haque-bnsm}. 

The scenario is somewhat different for the case of unequal mass binaries. At the initial stage, the composition of the two stars merging is different. The smaller star is likely to not have any quark component inside whereas the massive one is likely to have quite a fraction of QM at the core. The simulation results show that the GW amplitude has unique signals absent for equal mass binary mergers at the point of first contact. The lower the onset point, the more significant the phase difference (from unequal mass binary merger without QM seed in any stars). There is a spike in the phase difference at the point of first contact. The difference in the peak frequency and PSD increases with the decrease of onset density. The difference in the observational signals due to the appearance of quark matter at high density is more for unequal mass binary mergers \cite{haque-bnsm}.

\section{Summary and Conclusion}

In the article, we studied the GW signals coming from the PT of HM to QM at the core of isolated NSs and in binary NS mergers. The PT is more likely to happen during BNSM; however, even for cold, isolated NS, the PT can occur as the star slows down during its lifetime. As the stars slow down, the central density rises above the critical density beyond which QM can appear. Once it crosses the critical density, there is a quark seeding at the star's core. The PT transition in isolated NSs is considered a spin-down initiated process. As the quark core appears at the centre, there is a difference in the thermodynamic variable across the boundary of the quark core. If the difference is significant, the discontinuity gives rise to shock discontinuity which propagates from the core to the outer surface.

The shock-induced PT is thought to be a two-step process. In the 1st step as the shock moves outwards, it deconfines HM to quark matter. The deconfinement happens till the point where QM is more stable than HM. The deconfinement happens almost instantaneously, and it propagates along with the shock. Inside the star, the shock velocity is very high (almost some fraction of the speed of light), and the shock front crosses the star within a few tens of microseconds. As the shock propagates out, it changes the star's internal structure and thus, the quadrupole moment of the star changes. This gives rise to a burst type of GW signal with a significant amplitude, but the frequency is a few hundred kHz. In the meantime, a second front starts from the star's centre, which converts the deconfined 2-flavour QM to stable 3-flavour QM, converting excess down to strange quarks. This combustion front is relatively slower than the first deconfinement front as the weak interaction timescale, and diffusion is significant compared to the strong interaction timescale. The velocity of the front is at least two to three orders slower than the first front, and thus the combustion takes about a few tens of milliseconds. The GW amplitude associated with the combustion front is also four to five orders smaller than the first one; however, the frequency is in the few $ 100$ Hz. PT from isolated NSs is also associated with neutrino emission and deposition of energy similar to SGRB. Also, the tilt angle of the star changes suddenly, which can infer for the sudden appearance (or disappearance) of pulsars.

On the other hand, post-merger signals from BNSM can also give some information about the PT happening at high density. After the merging, it is expected that there will be a density rise in the post-merger product, and if it crosses the critical density, QM can appear at the core of HYS after the merging. With general relativistic simulation involving numerical relativity, one can simulate BNSM and the state after the merging. IF QM appears after the merging, there can be significant amplitude and phase change in the GW signal compared to results where QM does not appear. It is also seen that the difference in the GW signals is more significant for BNSM of unequal star mergers.

All the signals discussed above are challenging to detect with present-day detectors (like LIGO and VIRGO) but are expected to be detected with the next generations of detectors. Also, dedicated detectors for high-frequency range can also be very helpful in separating such signals. Any detection of such signals will confirm the presence of QM at the NS interior. In contrast, the absence of such signals can hint at smooth crossover transitions even at high density.

\bibliography{references}

\end{document}